\documentclass[journal,twoside]{IEEEtran}

\usepackage{chemarrow}
\usepackage{CJK}
\usepackage{fancybox}
\usepackage{times}
\usepackage{amsmath}
\usepackage{amssymb}
\usepackage{graphicx}
\usepackage{subfigure}
\usepackage{bm}
\usepackage{cite}

\usepackage{array}
\usepackage[linesnumbered,ruled]{algorithm2e}

\usepackage{algorithmic}

\usepackage{amsfonts,amssymb,amsmath,bm}
\usepackage{graphicx,subfigure}
\usepackage{cite,color}
\usepackage{booktabs}
\usepackage{threeparttable}
\usepackage{mathtools}
\usepackage{verbatim}

\begin{document}

\title{Energy Management and Cross Layer Optimization for Wireless Sensor Network Powered by Heterogeneous Energy Sources}

\author{Weiqiang~Xu,
Yushu Zhang,
Qingjiang Shi, 
Xiaodong Wang

\thanks{

Weiqiang Xu, Yushu Zhang, and Qingjiang Shi are with School of Information Science \& Technology, Zhejiang Sci-Tech University, Hangzhou, 310018, P. R. China. Email: wqxu@zstu.edu.cn.

Xiaodong Wang is with the Department of Electrical Engineering, Columbia University, New York,
NY, 10027, USA. Email: wangx@ee.columbia.edu.

Acknowledgments: We are very grateful to
Prof. Michael J. Neely at the University of Southern California,
Dr. Cristiano Tapparello at University of Padova,
and Prof. Osvaldo Simeone at New Jersey Institute of Technology,
 for very helpful discussions.}
}

\maketitle

\begin{abstract}
Recently, utilizing renewable
energy for wireless system has attracted extensive attention.
However, due to the instable energy supply and the limited battery capacity, renewable
energy cannot guarantee to provide the perpetual operation for wireless sensor networks (WSN).
The coexistence of renewable
energy and electricity grid
is expected as a promising energy supply manner to
remain function of WSN for a potentially infinite lifetime.
In this paper, we propose a new system model suitable for WSN,
taking into account multiple energy consumptions due to sensing, transmission and reception, heterogeneous energy supplies from renewable
energy, electricity grid and mixed energy,
and multi-dimension stochastic natures due to energy harvesting profile, electricity price and channel condition. A discrete-time stochastic cross-layer optimization problem is formulated to achieve the optimal trade-off between the time-average rate utility and electricity cost subject to the data and energy queuing stability constraints.
The Lyapunov drift-plus-penalty with perturbation technique and block coordinate descent method is applied to obtain a fully distributed and
low-complexity cross-layer algorithm only requiring knowledge of the instantaneous system state.
The explicit trade-off between the optimization objective and queue backlog is theoretically proven. Finally, through the extensive simulations, the theoretic claims are verified, and the impacts of a variety of system parameters on overall objective, rate utility and electricity cost are investigated.

\end{abstract}

\begin{IEEEkeywords}
Wireless sensor networks, energy harvesting, electricity grid, heterogeneous energy, cross-layer optimization,  Lyapunov optimization, drift-plus-penalty, block coordinate descent.
\end{IEEEkeywords}


\section{Introduction}


Wireless sensor network (WSN) consist of a lot of spatially distributed autonomous sensor
nodes with limited energy, computation and sensing capabilities, to
monitor physical phenomena, and to cooperatively
transmit their data to a sink. WSN
have a variety of potential applications,
ranging from
multimedia surveillance, environmental
monitoring,  and advanced health care delivery to industrial process control.
Traditionally,
sensor nodes are powered by a non-rechargeable battery with limited energy storage capacities.
However, a lot of applications are expected to operate over a virtually infinite lifetime.
The energy scarcity represents one of the major limitations of WSN. Indeed, the post-deployment replacement of the sensors batteries is generally not practical or even impossible.
Thus, a variety of hardware optimizations, energy management policies and energy-aware network protocols have been proposed to carefully manage the limited energy
resources and thus to prolong the lifetime of a WSN\cite{Akyildiz_Survey,Weiqiang_Xu2014TWC2,Weiqiang_Xu2010JSAC}.

Recent advances in hardware design have
made energy harvesting (EH) technology possibly applied in wireless systems.
Sensor node equipped with EH device replenishes energy from renewable sources with a potentially infinite amount of available energy\cite{Sudevalayam_Survey,He_Chen,Zhang2014}.
Since EH technology is essentially different from the traditional non-rechargeable battery,
a new energy management policy is expected to well-match with the energy
replenishment process.
As such, a great deal of research efforts have been devoted to investigate the
energy management and data transmission in the EH powered scenario.
Some efforts proposed the optimal schemes  to achieve
 the maximum throughput, the minimum transmission completion time, and/or  the minimum information distortion
for a single EH node with finite or infinite data buffer and finite or infinite battery capacity \cite{Sharma2010,Tutuncuoglu,OzelJSAC2011,Srivastava2013,Castiglione2012, ZhangTWC,MaoTVT2014}.
However, for wireless multihop network powered by EH, different nodes may have quite different workload requirements and
available energy sources.
Due to the fact that the network performance is tightly coupled with energy management policy and mechanisms at the physical, MAC, network, and transport layers,  a limited amount of works investigated the cross-layer optimization  schemes in \cite{Fan_Networking,Mao2012TAC,Chen2012INFOCOM,Sarkar2013,Xu_IJSN}.
In particular, some works of cross-layer optimization leveraged Lyapunov optimization techniques.
Gatzianas et al. in \cite{Gatzianas2010}
applied Lyapunov techniques to design
an online adaptive transmission scheme for wireless networks
with rechargeable battery to achieve total system utility maximization
and the data queue stability.
Huang et al. in \cite{Huang_Neely} applied Lyapunov optimization techniques with weight perturbation \cite{NeelyPerformance2011} to achieve
a close-to-optimal utility performance in finite energy buffer. The proposed technique obtains an explicit and controllable tradeoff between optimality gap and queue sizes.
 Similarly, by adopting perturbation-based Lyapunov techniques, Tapparello et al. in \cite{Tapparello2014}
proposed the joint optimization scheme of source coding and transmission to minimize the reconstruction distortion cost for the correlated sources measurement.
All the above-mentioned works showed that network-wide cross layer optimization
is helpful for achieving the performance gain.
However,
the works mentioned above are still not suitable to efficiently deal with
the energy scarcity limitation of WSN.
There are still several technical challenges, including:

\textbf{A. Multiple energy consumption} A sensor node is equipped with a sensing module for data measurements and processing, and a communication module for data transmission and data reception. Almost all of the works mentioned above only account for the energy consumed in data transmission.
Traditionally,
energy consumption is known to be dominated by the communication module. However, this is not always true.
In \cite{Margi2006}, it was shown that communication-related tasks were possibly
less energy consumption than intensive processing,
and data transmission is only a slight
more energy consumption than data reception.
There exists a very limited works in \cite{Castiglione2012}, \cite{MaoTVT2014}, \cite{Liu_Simeone} to investigate the problem of energy allocation accounting for the energy requirement of data transmission and sensing together, only suitable for a single EH nodes.
To the best of our knowledge, so far, almost no works, except\cite{Tapparello2014}, studied the joint energy allocation for communication module and sensing module together in the multihop scenario.

\textbf{B. Hybrid energy supply} Due to the low recharging rate and the time-varying profile
of the energy replenishment process, sensor nodes solely powered by harvested energy can not guarantee to provide reliable services for the perpetual operation.
They may currently be suitable only
for very-low duty cycle devices.
Other complementary stable energy
supplies should be required to remain
a perpetual operation for WSN.
As the electricity grid (EG) is capable of providing persistent
power input, the coexistence of renewable
energy and electricity grid is considered as a promising technology to
tackle the problem of simultaneously guaranteeing the network operation and minimizing the electricity grid energy consumption, which had been confirmed in single-hop wireless system \cite{Huang_Lau}\cite{Gong_Niu}.
However, as far as we know,
no prior work addressed to cross-layer optimization for WSN powered by heterogeneous energy sources in multihop scenario.

\textbf{ C. Fully distributed implementation} In WSN, the entire system state is characterized by channel condition, energy harvesting profile, electricity price, data queue size and energy queue size.
Therefore, the centralized solution requiring the entire system state will lead to heavy signaling overhead and high computational complexity in the central optimizer.
Furthermore, this information about the entire system state may
be hard to obtain or even unattainable in practical implementation.
It is desirable to have the distributed optimization based on local information only.
Some existing works designed the partly distributed optimization solution in WSN powered by EH. For instance, in \cite{Huang_Neely}\cite{Tapparello2014},
the power allocation problem still requires centralized optimization.
However,
a partly distributed optimization solution is still impractical, or too costly in large-scale networks.
Fully distributed optimization solution is particularly attractive.
This motivates us to address a novel energy management and cross-layer optimization for WSN powered by heterogeneous energy sources.
The key contributions of this paper are summarized as follows:

(1) We  propose a more realistic energy consumption
model, which takes the energy consumption of sensing, transmission and reception into account.
We propose a new heterogenous energy supply model suitable for the node powered by renewable energy or/and electricity grid.
We also consider the multi-dimension stochastic natures from channel condition, energy harvesting profile and electricity price.
For such a model, we formulate a discrete-time stochastic cross-layer optimization problem
in WSN with the goal of  maximizing the time-average utility of the source rate and the time-average cost of energy consumption in electricity grid subject to the data and energy queuing stability constraints.



(2) To obtain a distributed and
low-complexity solution,
we apply the Lyapunov drift-plus-penalty with perturbation
technique \cite{NeelyPerformance2011} to transform the stochastic optimization problem into a series of iterations of the deterministic optimization problems.
Furthermore,
by exploiting the special structure, we design a fully distributed algorithm---Energy mAnagement and croSs laYer Optimization (EASYO) which decomposes the deterministic optimization problem into
the energy management (including energy harvesting and energy purchasing), source rate control (implicitly including energy allocation for sensing/processing), routing selection (implicitly including energy allocation for data reception), session scheduling and transmission power allocation.
EASYO is a fully distributed algorithm
which makes greedy decisions at each time slot without requiring
any statistical knowledge of the channel state, of the harvestable energy state and of the electricity price state.
Note that our proposed fully distributed algorithm is different from the cross-layer optimization algorithms in \cite{Huang_Neely}\cite{Tapparello2014},
where the transmission power allocation problem is optimized in the centralized manner,
leading to the huge challenging in practical implementation.

(3)  We analyze the performance of the proposed distributed algorithm EASYO, and show that a control parameter $V$  enables an explicit trade-off between the average objective value and queue backlog. Specifically, EASYO can achieve a time average objective value that is within ${\cal {O}}(1/V)$ of the optimal objective for any $V >0$, while ensuring that the average queue backlog is ${\cal  {O}}(V)$.
Finally, through the extensive simulations, the theoretic claims are verified, and the impacts of a variety of system parameters on overall objective, rate utility and electricity cost are investigated.




Throughout this paper, we use the following notations.
The probability of
an event $A$ is denoted by Pr$(A)$. For a random variable
$X$, its expected value is denoted by $\mathbb{E}[X]$ and its expected
value conditioned on event $A$ is denoted by $\mathbb{E}[X|{A}]$.
The indicator function for an event $A$ is denoted by $\textbf{1}_{A}$; it equals
1 if $A$ occurs and is 0 otherwise. ${{{[x]}^ + } = \max (x,0)}$.





The remainder of the paper is organized as follows. In Section II, we give the system model and problem formulation.
In Section III, we present
the distributed cross-layer optimization algorithm.
In Section IV, we present the performance analysis of
our proposed algorithm.
Simulation results are given in Section V. Concluding remarks are provided in Section VI.

\section{System Model and Problem Formulation}
\label{sec:model}

\begin{figure}
\centering
\includegraphics[scale=0.6,bb=31 261 400 530]{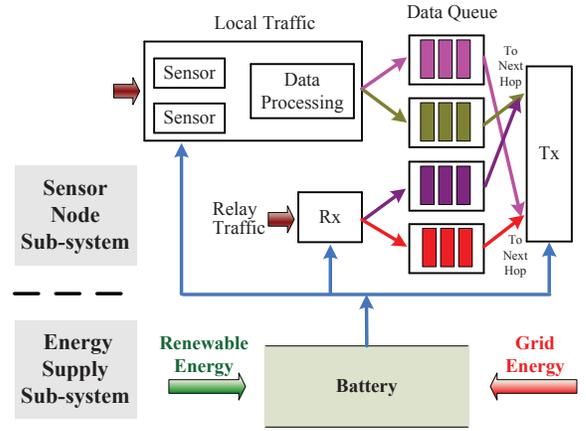}
\centering
\caption{Diagram of a single node system.}
\centering
\label{fig:NodeComposition}
\centering
\end{figure}

We consider a general interconnected multi-hop WSN that perfect CDMA-based
medium access, and operates over time slots $t \in {\cal T}=\left\{ {0,1,2, \ldots } \right\}$. WSN is modeled by a direct graph
${\cal G} = \left( {{\cal N},{\cal L}} \right)$.
${\cal N}  = {{\cal N}_H} \cup {{\cal N}_G} \cup {{\cal N}_M} = \left\{ {1,2,3, \ldots ,N} \right\}$ denotes the set of sensor nodes in the network, ${\cal N}_H$ is the set of nodes powered by EH, called EH nodes, ${\cal N}_G$ is the set of nodes powered by EG, called EG nodes, and ${\cal N}_M$ is the set of Mixed energy (ME) nodes powered by both EH and EG, respectively. ${\cal N}_s \subset {\cal N}$ denotes the set of all source nodes which measure the information source(s). Each source node $n \in {\cal N}_s$ has multiple sensor interfaces, such that it can measure multiple information sources ${\cal F}_n = \left\{ {1,2,3, \ldots ,F_n} \right\}$ at the same time \footnote{
In the following, we use the terms information source, flow and session
interchangeably. }. We use ${\cal F} = \bigcup \limits_{n \in {\cal N}_s}{{\cal F}_n}=\{1,2,\ldots, F\}$ to denote the set of all information sources in the network.
The source node  transmits the data to the corresponding destination node through multi-hop routing.
${\cal L}{\rm{ = }}\left\{ {(n,m),n,m \in {\cal N}} \right\}\,$ represents the set of communication links. ${\cal O}\left( n \right)$ denotes the set of nodes $m$ with $(n, m) \in \cal L$,
and ${\cal I}\left( n \right)$ denotes the set of nodes $m$ with $(m, n) \in \cal L$.
Fig. \ref{fig:NodeComposition} describes the composition of a single node system. The key notations of our system model are shown in TABLE I.
\begin{table}[t]
   \caption{\label{tab:table1}Summary of key notations}
\begin{tabular}{lll}
  \toprule
  Notation & Description\\
  \midrule
  ${\cal N}$ & The set of sensor nodes\\
  ${{\cal N}_H}$ & The set of nodes powered by EH\\
  ${{\cal N}_G}$ & The set of nodes powered by EG\\
  ${{\cal N}_M}$ & The set of nodes powered by both EH and EG\\
  ${{\cal N}_s}$ & The set of source nodes\\
  ${{\cal N}_d}$ & The set of destination nodes\\
  ${\cal F}$ & The set of all information sessions\\
  ${\cal L}$ & The set of communication links\\
  ${\cal O}\left( n \right)$ & The set of nodes $m$ with $(n, m) \in \cal L$\\
  ${\cal I}\left( n \right)$ & The set of nodes $m$ with $(m, n) \in \cal L$\\
  ${r_f}$ & The source rate of $f$-th information session\\
  ${p_{nb}^T}$ & The transmission power of link $\left( {n,b} \right)$\\
  ${\widehat{\gamma} _{nb}}$  & The signal to interference plus noise ratio (SINR) of \\
  &link $\left( {n,b} \right)$\\
  $x _{{nb}}^f\left( t \right)$  & The data transmission rate of the session $f$ over \\
  &link $\left( {n,b} \right)$\\
  ${\tilde{C }_{nb}}$ & The link capacity of link $\left( {n,b} \right)$\\
  ${\tilde{P}_f^S}$ & The energy consumption per data of the $f$-th session\\
  &for data sensening/processing\\
  ${\tilde P_n^R}$ &The energy consumed when node $n$ receiving one unit\\
  & data from the neighbor nodes in the network\\
  $P_n^{G}$ &The cost per unit of electricity drawn from the electricity\\
  & grid at node $n \in {\cal N}_G \cup {\cal N}_M$\\
  $p_n^{Total}$  &The total energy consumption of node $n$\\
  $\theta_n^E$  & The limited battery capacity of node $n$.\\
  $e_n$ &The harvested energy at node $n$\\
  $g_n$ &The energy  supplied by the electricity grid at node $n$\\
  ${S}_{nb}^C$ & The channel state of link $\left( {n,b} \right)$\\
 ${\bm S}^H$ & The harvestable energy state of node $n \in {{{\cal N}_H} \cup {{\cal N}_M}}$\\
  $S_n^G$ & The electricity price state at node $n$\\
  ${h_n}$ &The available amount of harvesting energy at node $n$\\
  ${{E_n}}\;$ &The energy queue size for $\,n \in {\cal N}\,$\\
  $Q_{n}^f\;$ & The data backlog of the $f$-th session at node $n$\\
  \bottomrule
\end{tabular}
\end{table}


\subsection{Source Rate and Utility}

At time slot $\,t\,$, the node $n$ measures $F_n$ independent
parallel information sources ${\cal F}_n$. The measured samples of the session $f \in {\cal F}_n$
is compressed with rate ${r_f}(t)$ before putting into the data queue\footnote{We measure time in unit size ¡°slots,¡± for simplicity, and thus we suppress the implicit multiplication by 1 slot when converting between data rate and data amount.},
where ${r_f}(t)$ denotes the source rate of the session $f$ at time slot $\,t\,$.
We assume that
\begin{equation}\label{source_rate_constraint}
 0 \le {r_f}(t) \le r_f^{\max }{\rm{,}}\forall f \in {\cal F}
\end{equation}
where $r_f^{\max }\le {R_{\max }}$ for all $f$ with some finite $R_{\max}$ at all time.
We assume that each session $f$  is associated
with a utility function $U_f({r_f(t)})$, which is
increasing, continuously differentiable and strictly concave in
$r_f(t)$  with a bounded first derivative and $U_f (0) = 0$. We use
$\beta_U^f(t)$ to denote the maximal first-order derivative of $U_f(r_f(t))$ w.r.t. $r_f(t)$,
denote $\beta_U=\max_{f \in {\cal F}, t\in \cal{T}}\;\;\beta_U^f(t)$.

\subsection{Data Transmission}

We assume that the links in the network may interfere with each other when they  transmit data simultaneously.
We define ${{\bm p}^T}\left( t \right){\rm{ = }}\left( {{p_{nb}^T}\left( t \right),\left( {n,b} \right) \in {\cal L}} \right)$  as the transmission power allocation matrix for data transmission at slot $t$, where ${p_{nb}^T}(t)$  is the transmission power allocated of link $\left( {n,b} \right)$, and then the following inequality should be satisfied:
\begin{equation}\label{energy constraint}
 0 \le \sum\limits_{b \in {\cal O}\left( n \right)} {{p_{nb}^T}(t)}  \le {P_n^{\max }}, n \in \cal N.
\end{equation}
where ${P_n^{\max }}$ is a finite constant to denote the maximal transmission power limitation at node $n$.

We use ${{\widehat{\gamma} } _{nb}}(t)$ to denote the signal to interference plus noise ratio (SINR) of link $\left( {n,b} \right)$:
\begin{eqnarray*}
{\widehat{\gamma} _{nb}}(t) &\buildrel\Delta \over =& {\widehat{\gamma} _{nb}}\left( {{{\bm p}^T}(t),{\bm S}^C(t)} \right)\\
&=&\frac{{{S _{nb}^C(t)}{p_{nb}^T}\left( t \right)}}{{{N_0^{b}} +
\sum_{a \in {\cal J}_{n,b}}\sum_{\left( {a,m} \right) \in {\cal L}}  {{S _{ab}^C(t)}{p_{am}^T}\left( t \right)} }},\notag
\end{eqnarray*}
where ${N_0^{b}}$ is the noise spectral density at node $b$, and ${S _{nb}^C}(t)$ represents the link fading coefficient from $n$ to $b$ at the slot $t$.
${\cal J}_{n,b}$ is the set of nodes whose transmission may interfere with the receiver of link $(n,b)$, excluding node $n$.
We assume that
${S _{nb}^C}(t)$ may be time varying and independent and identically distributed (i.i.d.) at every slot.
Denote ${\bm S}^C(t)=\left({S}_{nb}^C(t), (n, b) \in \cal L\right)$ as
the network channel state matrix, taking non-negative values from a
finite but arbitrarily large set ${\cal S}^C$.

The link capacity is defined as
\begin{equation*}
{\tilde{C }_{nb}}(t)= \log \left( {1 + K_{nb}{\widehat{\gamma}  _{nb}}\left( t \right)} \right).
\end{equation*}
Here, $K_{nb}$ denotes the processing gain of the CDMA system.
Note that the dependence of ${\tilde{C }_{nb}}(t) $ on ${{\bm p}^T}(t)$
and ${{\bm S}^C(t)}$ is implicit for notational convenience.
Let $x _{{nb}}^f\left( t \right)$  denote the data transmission rate of the session $f$ over link $\left( {n,b} \right)$, $b \in {\cal O}\left( n \right)$. Because of the total rates of all sessions cannot exceed the link capacity, so, $ 0 \le \sum\limits_{f \in {\cal F}} {x _{{nb}}^f\left( t \right)} \le {\tilde{C }_{nb}}\left( t\right), \forall n \in {\cal N}, \forall b \in {\cal O}\left( n \right)$.
Due to the fact that $K_{nb}$  is typically very large in CDMA networks,
$C _{nb}(t)=\log {\gamma}  _{nb}(t)$ is a good approximation of $\tilde{C } _{nb}(t)=\log(1+{\gamma}  _{nb}(t))$, where ${\gamma} _{nb}(t)=K_{nb}\widehat{\gamma} _{nb}(t)$. Thus, we make a stricter bound by the following constraint:
\begin{equation}\label{link_rate constraint}
0 \le \sum\limits_{f \in {\cal F}} {x _{{nb}}^f\left( t \right)} \le C _{nb}(t), \forall n \in {\cal N}, \forall b \in {\cal O}\left( n \right).
\end{equation}
Without loss of generality, we assume that for all time over all links under any power
allocation matrix and any channel state, there exists some finite constant $X_{\max}$.

\subsection{Energy Consumption Model}\label{energy_supply_sub-system1}
At every time slot $t$, each node $n$ allocates power\footnote{We measure time in unit size ¡°slots,¡± for simplicity, and thus we suppress the implicit multiplication by 1 slot when converting  between power and energy.} to accomplish its tasks, including data sensening/processing, data transmission and data reception.
We define a function $p_f^S(r_f(t))$ to denote the energy consumption of sensing/processing module for acquiring the data at a particular
rate $r_f(t)$ of the session $f$ at node $n$.
Inspired by \cite{Tapparello2014}, we also assume a linear relationship between the rate $r_f(t)$ and $p_f^S(r_f(t))$, i.e.,
$p_f^S(r_f(t))=\tilde{P}_f^Sr_f(t)$,
where ${\tilde{P}_f^S}$ denotes the energy consumption per data of the $f$-th session
for data sensening/processing. Thus,
the total energy consumption $p_n^{Total}\left( t \right)$ of node $n$ at slot $t$ is:
\begin{eqnarray}\label{EnergyConsumption}
p_n^{Total}\left( t \right) &\buildrel\Delta \over =&\sum\limits_{f \in {{\cal F}_n}} {\tilde{P}_f^S}{r _f}\left( t \right) \nonumber\\
 &+& \sum\limits_{b \in {\cal O}\left( n \right)} {p_{nb}^T\left( t \right)} +\tilde P_n^R \sum\limits_{a \in {\cal I}\left( n \right)} {\sum\limits_{f \in {\cal F}} {x _{{an}}^f\left( t \right)} }
\end{eqnarray}
where ${\tilde P_n^R}$ is the energy consumed when node $n$ receiving one unit data from the neighbor nodes in the network.


\subsection{Energy Supply Model}\label{energy_supply_sub-system2}

First, we describe the energy supply model of ME node shown in Fig. \ref{fig:NodeComposition}.
Each ME node is equipped with a battery having the limited capacity $\theta_n^E$.
As depicted in Fig. \ref{fig:NodeComposition}, the harvested energy $e_n(t)$ at time $t$ for
ME node $n$ is stored in the battery.
On the other hand, the energy supplied by the electricity grid at time $t$ for
ME node $n$ is denoted with $g_n(t)$.
Different from the ME node, the EH node only stores the harvested energy $e_n(t)$ and the EG node only stores the energy $g_n(t)$ supplied by the electricity grid.

We assume each  $n$ knows its own current energy
availability ${{E_n}}\left( t \right)\;$ denoting the energy queue size for $\,n \in {\cal N}\,$ at time slot $t$.
We define ${\bm E}\left( t \right) = \left( {{{E_n}}\left( t \right),n \in {\cal N}} \right)$ over time slots $t \in \cal T$ as the vector of the energy queue sizes.
The energy queuing dynamic equation is
\begin{eqnarray}\label{EMqueue}
E_n\left( {t + 1} \right) &=& E_n\left( t \right) + \textbf{1}_{n \in {{{\cal N}_H} \cup {{\cal N}_M}}}{e_n}\left( t \right)\nonumber\\
&+& \textbf{1}_{n \in {{{\cal N}_G} \cup {{\cal N}_M}}}{g_n}\left( t \right) - p_n^{Total}\left( t \right)
\end{eqnarray}
with $E_n\left( 0 \right) =0$.
At any time slot $t$, the total energy consumption at  node $n$ must satisfy the following \emph{energy-availability} constraint:
\begin{equation}\label{EMenergy_availability}
 {{E_n}}\left( t \right)\geq p_n^{Total}\left( t \right),\;\; \forall n \in {{\cal N}}.
\end{equation}

At any time slot $t$, the total energy volume stored in battery is limited by the battery capacity, thus the following inequality must be satisfied
\begin{equation}\label{limitedcapacity}
{{E_n}}(t)+{ \textbf{1}_{n \in {{{\cal N}_H} \cup {{\cal N}_M}}}{{e_n}\left( t \right)} + {\textbf{1}_{n \in {{{\cal N}_G} \cup {{\cal N}_M}}}{g_n}\left( t \right)} } \le{ \theta _{n}^{{E}}}
\end{equation}
%
%
%
%
%
%
We assume the available amount of harvesting energy at slot $t$  is ${h_n}\left( t \right)$
 with ${h_n}\left( t \right) \le {h_{\max }}$ for all $t$.
The amount of actually harvested energy ${e_n}\left( t \right)$ at slot $t$, should satisfy
 \begin{equation}\label{harvesting_decision}
0 \leq {e_n}\left( t \right) \leq  {h_n}(t), \forall n \in {{{\cal N}_H} \cup {{\cal N}_M}},
 \end{equation}
where ${h_n}\left( t \right)$  is randomly varying over time slots in an i.i.d. fashion according
to a potentially unknown distribution and taking non-negative values from a
finite but arbitrarily large set ${\cal S}^H$.
We define the harvestable energy state ${\bm S}^H\left( t \right) = \left( {{h_n}\left( t \right), n \in {{{\cal N}_H} \cup {{\cal N}_M}} }\right)$.

The energy supplied by the electricity grid ${g_n}\left( t \right)\;$  of the battery of node $n$ at slot $t$ should satisfy:
 \begin{equation}\label{charging_rate constraint}
 0\le {g_n}\left( t \right) \le g_n^{\max }, \forall n \in {{{\cal N}_G} \cup {{\cal N}_M}},
 \end{equation}
 with some finite $g_n^{\max }$.


\subsection{Electricity Price Model}
The cost per unit of electricity drawn from the electricity grid at node $n \in {\cal N}_G \cup {\cal N}_M$ at slot $t$
is denoted by $P_n^{G}(t)$. In general, it may depend on both $g_n(t)$,
the total amount of electricity from the electricity grid at slot $t$, and an electricity price
state variable $S_n^G(t)$, which represents such as both spatial and temporal variations, etc. For example, the per
unit electricity cost may be higher during daytime, and lower at late night.
We assume that $S_n^G(t)$ is randomly varying over time slots in an i.i.d. fashion according
to a potentially unknown distribution and taking non-negative values from a
finite but arbitrarily large set ${\cal S}^G$.
Denote ${\bm S}^{G}(t)=\left(S_n^{G}(t), n \in {{{\cal N}_G} \cup {{\cal N}_M}}\right)$ as the electricity price vector.
Similarly in \cite{Urgaonkar_Neely}, we assume that
$P_n^{G}(t)$ is a function of both $S_n^G(t)$ and $g_n(t)$, i.e.,
\begin{equation*}
P_n^{G}(t)= P_n^{G}(S_n^G(t), g_n(t))
\end{equation*}
Note that the dependence of $P_n^{G}(t)$ on $S_n^G(t)$
and $g_n(t)$ is implicit for notational convenience in the sequel.
For each given $S_n^G(t)$, $P_n^{G}(t)$
is assumed to be a increasing and continuous convex function of $g_n(t)$.
Let $\beta_G^1$ and $\beta_G^2$ denote the maximum and minimum unit electricity price in any slot in any node, respectively.

\subsection{Data Queue Model}

For $f \in {\cal F}$ at node $n$, we use $Q_{n}^f\left( t \right)\;$ to denote the data backlog of the $f$-th session at time slot $t$.
We define ${\bm{Q}}\left( t \right) = \left( {Q_n^f\left( t \right),n \in {\cal N},f \in {\cal F}} \right)$ over time slots $t \in \cal T$ as the data queue backlog vector. Then the data queuing dynamic equation is
\begin{eqnarray}\label{data_queue}
Q_n^f(t{\rm{ + 1}}) &=& Q_n^f(t) - \sum\limits_{b \in {\cal O}\left( n \right)} {x _{{nb}}^f\left( t \right)} \nonumber\\
&+& \sum\limits_{a \in {\cal I}\left( n \right)} {x _{{an}}^f\left( t \right)}+{\textbf{1}_{f \in {{\cal F}_n}}}{r_f}\left( t \right).
\end{eqnarray}
with $Q_{n}^f\left( 0 \right) =0$.
In any time slot $t$, the total data output at  node $n$ must satisfy the following \emph{data-availability} constraint:
\begin{equation}\label{Data_availability}
0\leq \sum\limits_{b \in {\cal O}\left( n \right)} {x _{{nb}}^f\left( t \right)} \leq {{Q_n^f}}\left( t \right),\;\; \forall n \in {{\cal N}},f \in {{\cal F}}.
\end{equation}

To ensure the network is strongly stable, the following inequality must be satisfied:
\begin{equation}\label{qstable}
\mathop {\lim }\limits_{T \to \infty } \;\frac{1}{T}\sum\limits_{t = 0}^{T - 1} {\sum\limits_{n \in {\cal N}} {\sum\limits_{f \in {\cal F}} \mathbb{E}{\left\{ {Q_n^f\left( t \right)} \right\}} } }  < \infty.
\end{equation}

\subsection{Optimization Problem}\label{sec:opt}
The goal
is to design a full distributed algorithm that achieves the optimal trade-off between the time-average utility of
the source rate and the time-average cost of energy consumption
in electricity grid,
which subject to all of the
constraints described above. Specifically, we define
\begin{eqnarray}\label{object}
O\left( t \right) &=& {\varpi _1}\sum\limits_{f \in {\cal F}} {{U_f}\left( {{{r}_f}(t)} \right)}\\
&-&\,\left( {1 - {\varpi _1}} \right)\sum\limits_{n \in {{{\cal N}_G} \cup {{\cal N}_M}}} {\varpi _2}{P_n^{G}(t)}{g_n(t)}\notag
\end{eqnarray}
where $\varpi _1$ ($0\le  \varpi _1 \le 1$) is a weight parameter to combine the objective functions together into a single one, and $\varpi _2$ is a mapping parameter to ensure the objective functions at the same level.

Mathematically,
we will address the stochastic optimization problem \textbf{P1} as follows:
\begin{eqnarray}\label{opt}
\mathop {\mbox{maximize}}_{\{{\bm \chi}(t), t \in \cal T\}} &&\overline O = \mathop {\lim }\limits_{T \to \infty } \;\frac{1}{T}\sum\limits_{t = 0}^{T - 1} \mathbb{E}{\left\{ {O\left( t \right)} \right\}} \\
\mbox{subject to}&&
\eqref{source_rate_constraint},
\eqref{energy constraint},
\eqref{link_rate constraint},
\eqref{EMenergy_availability},
\eqref{limitedcapacity},
\eqref{harvesting_decision},
\eqref{charging_rate constraint},
\eqref{Data_availability},
\eqref{qstable}\notag
\end{eqnarray}
with
the queuing dynamics \eqref{EMqueue} for $\forall n \in {{\cal N}}$ and \eqref{data_queue} for $\forall n \in {\cal N}, \forall f \in {\cal F}$.

${\bm \chi }(t) \buildrel \Delta \over =({\bm e}(t),{\bm g}(t),{\bm{p}}^T(t),{\bm r}(t), {\bm x}(t))$ is the set of the optimal variables of the problem \textbf{P1}, where
${\bm e}(t)$, ${\bm g}(t)$, ${\bm{p}}^T(t)$, ${\bm r}(t)$, ${\bm x}(t)$ are the vector of ${e_n}(t)$, ${g_n}(t)$, ${p_{nb}^T}(t)$, ${r}_f(t)$, ${x}_{nb}^f(t)$, respectively.

\section{Distributed Cross-layer Optimization Algorithm: EASYO }\label{sec:alg}
In this section, we propose an Energy mAnagement and croSs laYer Optimization algorithm (EASYO) for the problem \textbf{P1}.
Based on the Lyapunov optimization with weight perturbation technique developed in  \cite{NeelyPerformance2011}, \cite{NeelyBook2010} and \cite{Neely2006Book}\footnote{The core idea of Lyapunov optimization theory can be shortly acquired from the two following linkage:

$\mathrm{http://en.wikipedia.org/wiki/Drift\_plus\_penalty}$

$\mathrm{http://en.wikipedia.org/wiki/Lyapunov\_optimization}$.}, EASYO will determine
the energy harvesting, and the energy purchasing, source rate control, energy allocation for sensing/processing, transmission and reception, routing and scheduling decisions.
EASYO is a fully distributed algorithm
which makes greedy decisions at each time slot without requiring
any statistical knowledge of the harvestable energy states, of the electricity price states and of the channel states.

\subsection{Lyapunov optimization}

First, we introduce the weight perturbation ${{\bm \theta}  ^{{E}}} = \left( {\theta _{n}^{{E}},n \in {{\cal N}}} \right)$. Note that the weight perturbation ${\theta _{n}^{{E}}}$ is the limited battery capacity of node $n$ defined in Section \ref{energy_supply_sub-system2}.
Then we define the network state at time slot $t$ as
\begin{equation}\label{networkstate}
{\bm{Z}}(t) \buildrel \Delta \over =({{\bm S}^C}(t), {{\bm S}^H}(t),{{\bm S}}^G(t),{\bm{Q}}(t), {\bm{E}}(t))
\end{equation}

Define the Lyapunov function as
\begin{equation}\label{Lyapunov_function}
L(t)=\frac{1}{2}\sum\limits_{n \in {\cal N}} {\sum\limits_{f \in {\cal F}} {{{\left( {Q_n^f(t) } \right)}^2}} }
+\frac{1}{2}\sum\limits_{n \in {{\cal N}}} {{{\left( {{{E_n}}\left( t \right) - \theta _{n}^{{E}}} \right)}^2}}.
\end{equation}
\textbf{Remark 3.1}
From (\ref{Lyapunov_function}), we can see that when minimizing the Lyapunov function $L(t)$, the energy queue backlog is pushed towards the corresponding perturbed variable value, and the data queue backlog is pushed towards zero, which ensure the strong network stability constraint \eqref{qstable}. Furthermore, as long as we choose appropriate perturbed variables according to \eqref{thetaeh} in Theorem 1 at the next section, the constraint \eqref{EMenergy_availability} will always be satisfied due to \eqref{ea1} in Theorem 1 at the next section. Thus, we can get rid of \eqref{qstable} and \eqref{EMenergy_availability} in the sequel.

Now define the drift-plus-penalty as
\begin{equation}\label{driftplus14}
{\Delta _V}\left( t \right) \buildrel \Delta \over =\mathbb{E} \left(L(t+1) - L(t)- VO(t)|{\bm{Z}}(t)\right)
\end{equation}
where $V$ is a non-negative weight,  which can be tuned to control $\overline{O}$ arbitrarily close to the optimum with a corresponding tradeoff in average queue size.

We have the following lemma regarding he upper bound of the drift-plus-penalty ${\Delta _V}\left( t \right)$:

\textbf{Lemma 1}: Under any feasible energy management, source rate control, transmission power allocation, routing and scheduling actions that can be implemented at time $t$, we have the upper bound of ${\Delta _V}\left( t \right)$ as follows
\begin{equation}\label{upperbound15}
 {\Delta _V}\left( t \right) \le B +\mathbb{E}\left( {\widetilde{\Delta} _V}\left( t \right)\left| {{\bm{Z}}(t)} \right.\right),
\end{equation}
where
\begin{equation}\label{B_definition}
B = NF{B_Q} + \sum\limits_{n \in {{\cal N}}} {B_E}
\end{equation}
with ${B_Q}{\rm{ = }}\frac{3}{2}l_{\max }^2{{X}} _{\max }^2 + {\left( {R_{\max }} \right)^{\rm{2}}}$, where ${l_{\max }}$ denotes the largest number of the outgoing/incoming links that any node in the network can have.
${B_E} =\frac{1}{2}{\left( {\textbf{1}_{n \in {{{\cal N}_H} \cup {{\cal N}_M}}}h_{\max }+\textbf{1}_{n \in {{{\cal N}_G} \cup {{\cal N}_M}}}g_n^{\max}} \right)^2}+\frac{1}{2}{\left( {P_{n,\max }^{Total}} \right)^2}$,
$P_{n,\max }^{Total} = \sum\limits_{f \in {{\cal F}_n}}{\tilde P_f^S{r_f^{\max }}} + P_n^{\max } + \tilde P_n^R{l_{\max }}{X_{\max }}$.

${\widetilde{\Delta} _V}\left( t \right)$ is given in \eqref{upperbound15B},
\begin{figure*}
 \begin{eqnarray}\label{upperbound15B}
{\widetilde{\Delta} _V}\left( t \right)&=& \sum\limits_{n \in {{\cal N}}}\left[\left( {{{E_n}}(t) - \theta _n^{{E}}} \right)\textbf{1}_{n \in {{{\cal N}_H} \cup {{\cal N}_M}}}{{e_n}\left( t \right)}+{\left( {D_n(t)+E_n\left( t \right) - \theta _n^e} \right) \textbf{1}_{n \in {{{\cal N}_G} \cup {{\cal N}_M}}}{g_n}\left( t \right)}\right]\\
 &-&\sum\limits_{n \in {\cal N}_s}  \sum\limits_{f \in {{\cal F}_n}} \left[V{\varpi _1}{U_f}\left( {{{r}_f}(t)} \right)-Q_n^f(t){{ r}_f}(t)+ A_{n}(t){\tilde{P}_f^S}{{ r}_f}(t) \right]  -  \sum\limits_{n \in {\cal N}} {\sum\limits_{b \in {\cal O}\left( n \right)} \left[\sum\limits_{f \in {\cal F}} W_{nb}^f(t)x _{nb}^f(t)+A_{n}(t)p_{nb}^T(t)\right] }\nonumber
\end{eqnarray}
\hrule
\end{figure*}
where
\begin{equation}\label{Definition of D}
 D_n(t)\buildrel \Delta \over ={V\,\left( {1 - {\varpi _1}} \right){\varpi_2}{P_n^{G}}(t)},
\end{equation}
\begin{equation}\label{Definition of A}
A_{n}(t) \buildrel \Delta \over = {{E_n}}(t) - \theta _n^{{E}},
\end{equation}
and
\begin{equation}\label{link_information_weight}
W_{nb}^{f}(t)\buildrel \Delta \over =Q_n^f(t) - Q_b^f(t)+A_{b}(t){\tilde{P}_b^R}.
\end{equation}
\textbf{Proof}: See Appendix A.

\subsection{Framework of EASYO}
We now present our algorithm EASYO. The main design principle
of EASYO is to minimize
the right hand side (RHS) of (\ref{upperbound15B}) subject to the constraints \eqref{source_rate_constraint},
\eqref{energy constraint},
\eqref{link_rate constraint},
\eqref{limitedcapacity},
\eqref{harvesting_decision},
\eqref{charging_rate constraint} and
\eqref{Data_availability}.

The framework of EASYO is described in Algorithm 1 summarized in TABLE II.
\begin{table}[htbp]
\centering
\caption{Algorithm 1: EASYO}
\begin{tabular}{|p{3in}|}
\hline
\begin{itemize}
\item [1] \; Initialization:
The perturbed variables ${{\bm \theta} ^{{E}}}$ and the penalty parameter $V$ is given.
\item [2] \; Repeat at each time slot $t \in \cal T$:
\item [3]  \quad Observe ${\bm Z} (t)$;
 \item [4]  \quad Choose the set ${\bm{\chi}^{*}(t)}$ of the optimal variables as the
optimal solution to the following optimization problem \textbf{P2}:
\begin{eqnarray}\label{P2}
\mathop {\mbox{minimize}}\limits_{\bm{\chi}(t)} && {\widetilde{\Delta} _V}\left( t \right)\notag\\
\mbox{subject to} &&\eqref{source_rate_constraint},
\eqref{energy constraint},
\eqref{link_rate constraint},
\eqref{limitedcapacity},
\eqref{harvesting_decision},
\eqref{charging_rate constraint},
\eqref{Data_availability}
\notag
\end{eqnarray}
\item [5]  \quad Update the energy queues and data queues according to \eqref{EMqueue} and \eqref{data_queue}, respectively.
\end{itemize}
\\
\hline
\end{tabular}
\end{table}

\textbf{Remark 3.2} Note that the algorithm EASYO only requires the knowledge
of the instant values of ${\bm Z} (t)$. It does not
require any knowledge of the statistics of these stochastic
processes. The remaining challenge is to solve the problem \textbf{P2}, which is
discussed below.

\subsection{Components of EASYO}
At each time slot $t$, after observing ${\bm Z} (t)$, all components of EASYO is iteratively implemented in distributed manner to cooperatively solve the problem \textbf{P2}.
Next, we describe each component of EASYO in detail.

(1) \textbf{Energy Management}
For each node $n \in {{\cal N}}$, combining the first term of the RHS of \eqref{upperbound15B} with the the constraint \eqref{limitedcapacity}, \eqref{harvesting_decision} and \eqref{charging_rate constraint}, we have the optimization problem of ${e_n}(t)$ and ${g_n}(t)$ as follows:
\begin{eqnarray}\label{Energy Harvesting and Battery Charging}
\mathop{\mbox{minimize}}\limits_{{e_n}(t), {g_n}(t)}&&\left( {{{E_n}}(t) - \theta _n^{{E}}} \right) \textbf{1}_{n \in {{{\cal N}_H} \cup {{\cal N}_M}}}{{e_n}\left( t \right)} \nonumber\\
&&+ {\left( {D_n(t)+E_n\left( t \right) - \theta _n^e} \right) \textbf{1}_{n \in {{{\cal N}_G} \cup {{\cal N}_M}}}{g_n}\left( t \right)} \nonumber\\
\mbox{subject to}&&0\le {e_n}(t) \le {h_n}(t)\label{eht}\\
                 &&0 \le {g_n}(t) \le g_n^{\max }\label{epc}\\
                 && { \textbf{1}_{n \in {{{\cal N}_H} \cup {{\cal N}_M}}}{{e_n}\left( t \right)} + {\textbf{1}_{n \in {{{\cal N}_G} \cup {{\cal N}_M}}}{g_n}\left( t \right)} }\nonumber\\
                 && \le{ \theta _{n}^{{E}}-{{E_n}}(t)}\label{limitedcapacity2}
\end{eqnarray}

\textbf{Remark 3.3}
Energy management component is composed of energy harvesting and energy purchasing. Furthermore, since $P_n^{G}(t)$
is increasing and continuous convex on $g_n(t)$ for each $S_n^G(t)$,
it is easy to verify that energy management component is a convex optimization problem in $({e_n}(t),{g_n}(t))$ , which can be solved efficiently.

\textbf{Remark 3.4}
From \eqref{limitedcapacity2}, we can see that all the incoming energy is stored if there is enough room in the energy buffer according to the limitation imposed by ${\theta _n^{{E}}}$, and otherwise it stores all the energy that it can, filling up the battery size of ${\theta _n^{{E}}}$. Hence, ${{E_n}}(t) <{\theta _n^{{E}}}$ for all $t$,
which means that
EASYO can be implemented with finite energy storage capacity ${\theta _n^{{E}}}$ at node $n \in \cal N$.

(2) \textbf{Source Rate Control}
For each session $f \in {\cal F}_n$ at source node $n \in {\cal N}_s$ , combining the second term of the RHS of \eqref{upperbound15B} with the constraint \eqref{source_rate_constraint}, we have the optimization problem of ${r_f}(t)$ as follows:
\begin{eqnarray}\label{SourceRate}
\mathop{\mbox{maximize}}\limits_{{r_f}(t)}&& V{\varpi _1}{U_f}\left( {{{r}_f}(t)} \right)-{\left(Q_n^f(t) - A_{n}(t){\tilde{P}_f^S}\right){{ r}_f}(t)} \notag\\
\mbox{subject to}&&0 \le {r_f}(t) \le r_f^{\max }
\end{eqnarray}
Let ${{{r}_f^*}}$ be the unique maximizer. By the Kuhn-Tucker theorem, ${{{r}_f^*}}$ is given by
\begin{equation}\label{SourceRateSolution}
{{{r}_f^*}}=\left[ {U_f^{' - 1}\left( Q_n^f(t) - A_{n}(t){\tilde{P}_f^S} \right)} \right]_0^{r_f^{\max }}
\end{equation}
where
$\left[ z \right]_a^b = \min \left\{ {\max \left\{ {z,a} \right\},b} \right\}$, $U_f^{' - 1}(\cdot)$ is the inverse of the derivative of $U_f(\cdot)$.

(3) \textbf{Joint Optimal Transmission Power Allocation, Routing and Scheduling}
Combining the third term of the RHS of \eqref{upperbound15B} with the constraints \eqref{energy constraint}, \eqref{link_rate constraint} and the data-availability constraint \eqref{Data_availability},
 we have the optimization problem of  ${\bm x}(t)$ and ${{\bm p}^T(t)}$ as follows:
\begin{flalign}\label{TPARS}
\mathop{\mbox{maximize}}\limits_{{\bm x}(t), {{\bm p}^T(t)}}&\;\;\sum\limits_{n \in {\cal N}} {\sum\limits_{b \in {\cal O}\left( n \right)} {\sum\limits_{f \in {\cal F}} W_{nb}^f(t)x _{nb}^f(t)+\left(A_{n}(t)p_{nb}^T(t)\right)}}  \notag\\
\mbox{subject to}&\;\;0\leq\sum\limits_{f \in {\cal F}} {x _{nb}^f}(t) \le {C_{nb}}(t),\;\forall n \in {\cal N},\forall b \in {\cal O}\left( n \right)\notag\\
 &\;\;0\leq\sum\limits_{b \in {\cal O}\left( n \right)} {p_{nb}^T(t)}  \le {P_n^{\max }}, \forall n \in {\cal N}\notag\\
 &\;\; 0\leq\sum\limits_{b \in {\cal O}\left( n \right)} {x _{nb}^f}(t) \le Q_n^f(t),  \forall f \in {\cal F}
\end{flalign}


Now, we will solve the optimization problem \eqref{TPARS}.
Define the weight of the session $f$ over link $(n,b)$ as:
\begin{equation}\label{weightnformation}
\tilde W_{nb}^f(t) \buildrel \Delta \over = {\left[ {W_{nb}^f(t) - \sigma} \right]^ + },
\end{equation}
where
\begin{equation}\label{gamma}
\sigma  = {l_{\max }}{X_{\max }} + {r_f^{\max }}
\end{equation}
denotes the data amount of the session $f$ which the node $n$ can receive at most at time slot $t$.


\textbf{Transmission Power Allocation Component}
For each node $n$, find any
${f_{nb}^*} \in \arg {\max _f}\;\tilde W_{nb}^f(t)$.
Define $\tilde W_{nb}^{^*}(t) = {{{\max }_f}\;\tilde W_{nb}^f(t)}$ as the corresponding optimal weight of link $\left( {n,b} \right)$.
Observe the current channel state ${\bm S}^C(t)$, and select the transmission powers
${\bm p}^{T*}$
by solving the following optimization problem:
\begin{eqnarray}\label{opeq}
\mathop{\mbox{maximize}}\limits_{{\bm p}^{T}}&&\sum\limits_{n \in {\cal N}} {\sum\limits_{b \in {\cal O}\left( n \right)} {\left( {\tilde W_{nb}^*(t){C_{nb}(t)} + {A_{n}}(t){p_{nb}}(t)} \right)} }  \nonumber\\
\mbox{subject to}&&0 \leq \sum\limits_{b \in {\cal O}\left( n \right)} {p_{nb}^T(t)}  \le {P_n^{\max }},\forall n \in {\cal N}
\end{eqnarray}

\textbf{Routing and Scheduling Component}
The data of session ${f_{nb}^*}$  is
selected for routing over link $(n, b)$ whenever
$\tilde W_{nb}^{{f^*}}(t) > 0$.
That is, if $\tilde W_{nb}^{{f^*}}(t) > 0$, set $x _{nb}^{{f_{nb}^*}}(t)={C_{nb}}\left({\bm p}^{T*}, {\bm S}(t) \right)$.

\textbf{Remark 3.5}:
If we set $ \sigma= 0$,
the joint transmission power allocation, routing and scheduling component
is to minimize the third term of the RHS in \eqref{upperbound15B}.
Inspired by \cite{NeelyBook2010} and \cite{Neely2006Book},
we set a non-zero $ \sigma$ in \eqref{weightnformation},  leading to a easy way to determine the upper bound of all queue sizes shown in Theorem 1.
Also the definition \eqref{gamma} of $ \sigma$ can ensure
the constraints \eqref{Data_availability}  will always be satisfied. The detailed proof will be given in Theorem 1. Thus, we can get rid of this constraint \eqref{Data_availability} in \eqref{opeq}.

\textbf{Remark 3.6}:
Our proposed EASYO is designed to minimize the RHS of \eqref{upperbound15B}.
Each component contributes to
minimizing the part of the RHS of \eqref{upperbound15B}.
Taking all components together, EASYO contributes to minimize the whole RHS of \eqref{upperbound15B}, and thus to
minimize ${\Delta _V}\left( t \right)$.
Because the whole RHS of \eqref{upperbound15B} incorporates the Lyapunov drift, EASYO is stable.
Meanwhile, since it also incorporates the objective of the problem \textbf{P1},  EASYO is optimal.


\textbf{Remark 3.7}:
The former two components of EASYO are computed in
closed form or numerically solved through a simple convex optimization problem, only based on the local information.
The unique challenge of distributed implementation of EASYO
is to distributedly solve the transmission power allocation problem \eqref{opeq}.
Next, we will develop the distributed algorithm.


\subsection{Distributed Implementation of Transmission Power Allocation}

After implementing a variable change ${\hat p_{nm}^T}(t) = \log \left( {{p_{nm}^T}(t)} \right)$, and taking the logarithm of both sides of the constraint in problem \eqref{opeq}, the problem \eqref{opeq} can be equivalently transformed into the problem \textbf{P3}
\begin{eqnarray}\label{transformed_power_allocation}
\mathop{\mbox{max}}\limits_{{\hat {\bm p}}^T(t)}&&\sum\limits_{n \in {\cal N}} {\sum\limits_{b \in {\cal O}\left( n \right)} {\left( {\tilde W_{nb}^*(t){\Psi _{nb}}\left( {{\bm{\hat p}^T}(t)} \right)  + {A_{n}}(t){e^{\hat p_{nb}^T(t)}}} \right)} }  \nonumber\\
\mbox{s.t.}&&\log \sum\limits_{b \in {\cal O}\left( n \right)}{e^{{\hat p}_{{nb}}^T(t)}}  - \log{P_n^{\max }} \le 0,\forall n \in {\cal N}
\end{eqnarray}
where ${\hat {\bm p}}^T(t)=(\hat{p}_{nb}^T(t), n \in {\cal N}, b \in {\cal O}\left( n \right))$, ${\Psi _{nb}}\left( {{\bm{\hat p}^T}(t)} \right)$ is defined in \eqref{Psi}.
\begin{flalign}\label{Psi}
&{\Psi _{nb}}\left( {{\bm{\hat p}^T}(t)} \right)&\\
& \buildrel \Delta \over= \log \left( {{\gamma _{nb}}(t)} \right)
 = \log {S _{nb}^C}(t) + \hat p_{nb}^T(t)&\nonumber \\
 &- \log \left( {N_0^b + \sum\limits_{a \in {\cal J}_{n,b}}\sum\limits_{\left( {a,m} \right) \in {\cal L}} {\exp \left( {\hat p_{am}^T(t) + \log {S _{ab}^C}(t)} \right)} } \right).&\nonumber
\end{flalign}
It is not difficult to prove that ${\Psi _{nb}}\left( {{\bm{\hat p}^T}} \right)$ is a strictly concave
function of a logarithmically transformed power vector $\bm{\hat p}^T(t)$\cite{Boyd_Convex}.
Due to \eqref{limitedcapacity} or \eqref{limitedcapacity2}, we have
${{E_n}}(t) \le \theta _n^{{E}}$,
 so $A_{n}(t) \le 0$,  thus ${A_{n}}(t){e^{\hat p_{{nb}}^T(t)}}$ is a strictly concave function of ${\hat p_{{nb}}^T(t)}$.
To sum up, the objective of \textbf{P3} is a strictly convex
in $\bm{\hat p}^T(t)$. Furthermore, since $\log \sum\limits_{b \in {\cal O}\left( n \right)} {{e^{\hat p_{{nb}}^T(t)}}}$  is a strictly concave
in $\bm{\hat p}^T(t)$,  \textbf{P3} is a strictly convex optimization problem, which has the global optimum.

To  distributedly solve \textbf{P3}, we propose a distributed iterative algorithm based
on block coordinate descent (BCD) method whereby,
at every iteration, a single block of variables is optimized while the remaining
blocks are held fixed. More specifically, at iteration $t_i$, which represents the $i$-th iteration at the time slot $t$, for each node $n \in {\cal N}$, the blocks ${\hat {\bm p}}_n^T=(\hat{p}_{nb}^T, b \in {\cal O}\left( n \right))$ are updated through
solving the following optimization problem \eqref{transformed_power_allocation1},
where ${\hat {\bm p}}_{-n}^T(t_i)=({\hat {\bm p}}_1^T(t_i), \cdots, {\hat {\bm p}}_{n-1}^T(t_i),{\hat {\bm p}}_{n+1}^T(t_i), \cdots, {\hat {\bm p}}_N^T(t_i))$ are held fixed.
\begin{eqnarray}\label{transformed_power_allocation1}
\mathop {\mbox{maximize}}_{{\hat {\bm p}}_n^T} &&\sum\limits_{n \in {\cal N}}\sum\limits_{b \in {\cal O}\left( n \right)} ( \tilde W_{nb}^*(t){\Psi _{nb}}\left( {{{\hat {\bm p}}_n^T},{{\hat {\bm p}}_{-n}^T}(t_i)} \right)  \nonumber\\
&&+ {A_{n}}(t){e^{\hat p_{nb}^T}} )   \nonumber\\
\mbox{subject to}&&\log \sum\limits_{b \in {\cal O}\left( n \right)}{e^{{\hat p}_{{nb}}^T}}  - \log{P_n^{\max }} \le 0.
\end{eqnarray}

The global rate of convergence for BCD-type algorithm has been studied extensively when
the block variables are updated in both
the classic Gauss-Seidel fashion and the randomized update rule\cite{ZQLuo}\cite{Beck}. Since the optimization problem \textbf{P3} is strongly convex in $\bm{\hat p}^T(t)$, our proposed BCD-based distributed iterative algorithm can converge to the global optimum of \textbf{P3}.

\section{Performance analysis}

Now, we analyze the performance of our proposed algorithm EASYO.
To start with, we assume that there exists $\delta>0$  such that
\begin{equation}\label{PowerFeature1}
{C_{nb}}\left( {{{\bm p}^T}\left( t \right),{\bm S}^C\left( t \right)} \right) \le \delta {p_{nb}^T}\left( t \right), \forall n \in {\cal N}, \forall b \in {\cal O}\left( n \right).
\end{equation}

\textbf{Theorem 1}:
Implementing the algorithm EASYO with any fixed
parameter $V>0$ for all time slots, we have the following
performance guarantees:

(\textbf{A}).
Suppose the initial data queues and the initial energy queues satisfy:
\begin{eqnarray}
0 \le Q_n^f\left( 0\right) &\le& Q_{\max},\quad n \in {\cal N}, f \in {\cal F}\label{Qbound0} \\
0 \le E_{n}\left( 0\right) &\le& \theta _{n}^{{E}},\quad n \in {{\cal N}}\label{ehu0}
\end{eqnarray}
where the queue upper bounds are given as follows:
\begin{equation}\label{Q_upbound}
Q_{\max}={\varpi _1}{\beta_U}V+ r_f^{\max },
\end{equation}
\begin{equation}\label{thetaeh}
\theta _n^{{E}}= \delta{\varpi _1}{\beta_U}V + P_{n,\max }^{Total}.
\end{equation}
Then, the data queues and the energy queues of all nodes for all time slots $t$ are always
bounded as
\begin{eqnarray}
0 \le Q_n^f\left( t \right) &\le& Q_{\max},\quad n \in {\cal N}, f \in {\cal F}\label{Qbound} \\
0 \le E_{n}\left( t \right) &\le& \theta _{n}^{{E}},\quad n \in {{\cal N}},\label{ehu}
\end{eqnarray}

(\textbf{B}).  The objective function value of the problem \textbf{P1} achieved by the proposed algorithm EASYO satisfies the bound
\begin{equation}\label{objective_value}
\overline{O} \ge {O^*} - \frac{{\tilde{B}}}{V}
\end{equation}
where ${O^*}$ is the optimal value of  the problem \textbf{P1},
and $\tilde{B}=B+NF\sigma l_{\max}X_{\max}$ .

(\textbf{C}).  When node $n \in {{\cal N}}$ allocates nonzero power for data sensing, data transmission and/or data reception, we have:
\begin{equation}\label{ea1}
E_{n}\left( t \right) \ge P_{n,\max }^{Total}, n \in {{\cal N}}.
\end{equation}

(\textbf{D}). For node $n \in {{\cal N}}$, when any data of the $f$-th session is transmitted to other node, we have:
 \begin{equation}\label{ea4}
Q_{n}^f\left( t \right) \ge l_{\max}{X}_{\max}.
\end{equation}
\textbf{Proof}: Please see Appendix B-E.

\textbf{Remark 4.1}:
Theorem 1 shows that a control parameter $V$  enables an explicit trade-off between the average objective value and queue backlog. Specifically, for any $V >0$,
the proposed distributed algorithm EASYO can achieve a time average objective that is within ${\cal {O}}(1/V)$ of the optimal objective shown in \eqref{objective_value}, while ensuring that the average data and energy queues have upper bounds of ${\cal  {O}}(V)$  shown in \eqref{Qbound} and \eqref{ehu}, respectively.
In the section \ref{Simulation}, the simulations will verify the theoretic claims.

\textbf{Remark 4.2}: The inequality \eqref{ea1} guarantees that the  \emph{energy-availability} constraint \eqref{EMenergy_availability} is satisfied for all nodes and all times. Similarly,
the inequality \eqref{ea4} ensures that
the \emph{data availability} constraint \eqref{Data_availability} is always satisfied.

\section{Simulation Results}\label{Simulation}

\begin{figure}
\centering
\includegraphics[scale=0.6,bb=120 480 475 755]{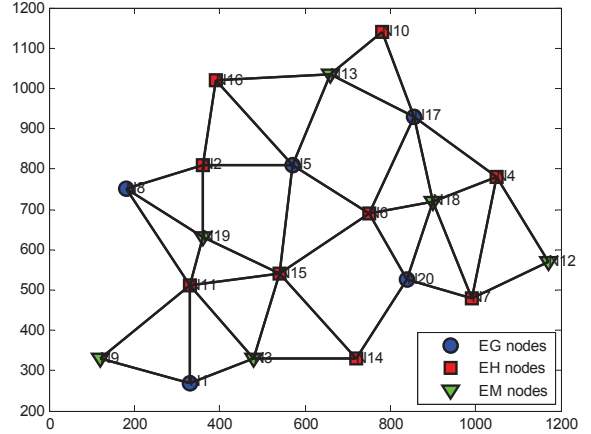}
\centering
\caption{Network topology.}
\centering
\label{fig:Topology}
\centering
\end{figure}

In this section, we provide the simulation results of the algorithm EASYO for the network scenario shown in Fig.\ref{fig:Topology}. In this scenario, we consider a multi-channel WSN with 20 nodes, 78 links, 6 sessions transmitted on 14 different channels. Throughout,
the form of the rate utility function is set as $U_f(r_f(t))=\log(1+r_f(t))$, so
${\beta _U}=1$.
The form of the electricity cost function is set as $P_n^{G}(t)=S_n^{G}(t)$.

Set several default values as follows:
$\delta=2$;
 $r_f^{\max }=3, \tilde{P}_f^S=0.1, \forall f \in {\cal F}$; $g_n^{\max }=2, \forall n \in {{\cal N}_G} \cup {{\cal N}_M}$; $X_{\max}=2, l_{\max}=6, P_n^{\max }=2, \tilde{P}_n^R=0.05, \forall n \in {\cal N}$; $N_0^b=5 \times{10^{ - 13}}$, $\varpi_1=0.6$, $\varpi_c=0.5$.
We set all the initial queue sizes to be zero.

The
channel-state matrix ${\bm S}^C(t)$ has independent entries that for every
link are uniformly distributed with interval $[S_{\min}^C, S_{\max}^C]\times d^{-4}$, $S_{\min}^C=0.9$, $S_{\max}^C=1.1$ as default values and $d$ denotes the distance between transmitter and receiver of the link, while the energy-harvesting vector ${\bm S}^H(t)$  has
independent entries that are uniformly distributed in $[0, h_{\max}]$,
with $h_{\max}=2$ as default value.
The electricity price vector ${\bm S}^{G}(t)$ has
independent entries that are uniformly distributed in $[S_{\min}^{G}, S_{\max}^{G}]$
with $S_{\min}^{G}=0.5$, $S_{\max}^{G} =1$ as default values, so
$\beta _G^1 = S_{\max}^{G}$, $\beta _G^2 = S_{\min}^{G}$.
All statistics of ${\bm S}^C(t)$,
${\bm S}^H(t)$ and ${\bm S}^{G}(t)$  are i.i.d. across time-slots.
%

\begin{figure}[t]
\centering
\includegraphics[width=2.8in]{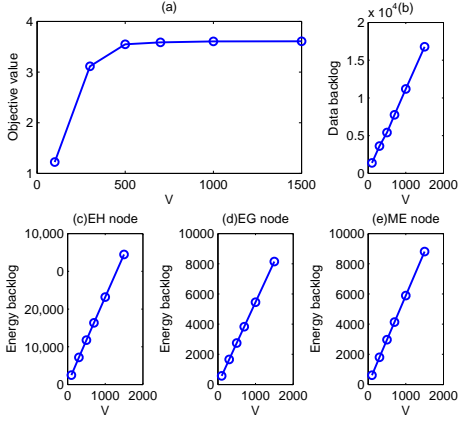}
\caption{Verification of Theorem 1.}
\label{fig_Verifying_theoretic_claims}
\end{figure}

We simulate $V$=[100,300,500,700,1000,1500].
In all simulations, the simulation time is $10^{5}$ time slots.
The simulation results are depicted in Fig. \ref{fig_Verifying_theoretic_claims}.
From Fig. \ref{fig_Verifying_theoretic_claims} (a), we see that as $V$ increases, the time average optimization objective value keep increasing and converge to very close to  the optimum. This confirms the results of \eqref{objective_value}.
From Fig. \ref{fig_Verifying_theoretic_claims}  (b), we see that as $V$ increases, the average data queue length
keeps increasing. From Fig. \ref{fig_Verifying_theoretic_claims}  (c)-(e), we observe that the battery queue size increases as $V$ increases.
A closer inspection of the results also reveals a linear increase
of the time average data and energy queue size with respect to $V$.
This shows a good match between the simulations and Theorem 1.

\begin{figure}[t]
\centering
\includegraphics[width=2.8in]{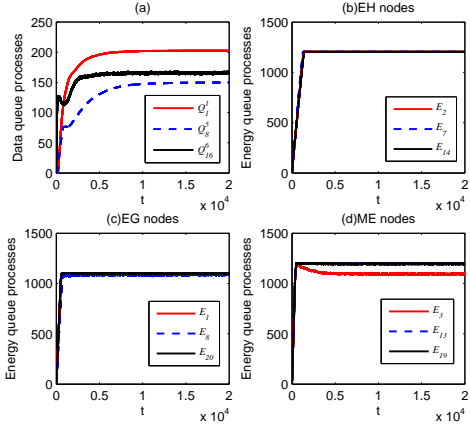}
\caption{Detailed verification of the queueing bounds.}
\label{fig_detailed_verification_queue_bound}
\end{figure}

\begin{figure}[t]
\centering
\includegraphics[width=3.2in]{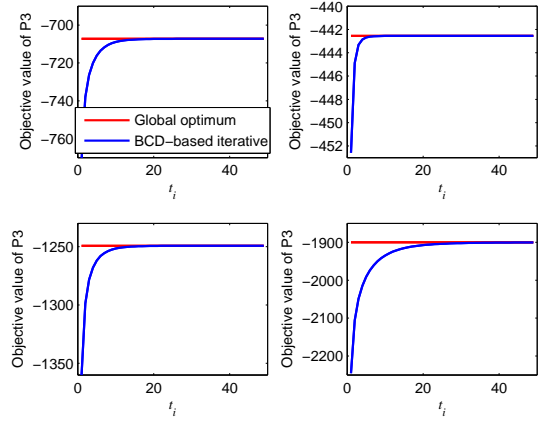}
\caption{Convergence of BCD-based distributed iterative algorithm.}
\label{fig:BCD_Conv}
\end{figure}

For better verification of the queueing bounds, we also present the data queue
process of node $1$ for session $1$, of node $8$ for session $5$ and of node $16$ for session $6$ under $V =1000$ in Fig \ref{fig_detailed_verification_queue_bound}(a), and the energy queue
processes for EH node $2, 7, 14$, for EG node $1, 8, 20$ and for ME node $3, 13, 19$ under $V =1000$ in Fig. \ref{fig_detailed_verification_queue_bound}(b)-(d), respectively.
It can be verify that all queue sizes can quickly converge with the upper bound given in Theorem 1.


Transmission power allocation problem \textbf{P3} is the most  complex component in our proposed EASYO.
We proposed a BCD-based distributed iterative algorithm to
solve the problem \textbf{P3}.
During the implementation of EASYO, we catch four different snapshots of the iterative procedure of BCD-based algorithm, shown in Fig. \ref{fig:BCD_Conv}.
From Fig. \ref{fig:BCD_Conv}, we can see that BCD-based algorithm can quickly converge to the global optimum. Thus, our proposed EASYO is a low-complexity distributed algorithm.

\begin{figure}[t]
\centering
\includegraphics[width=2.8in]{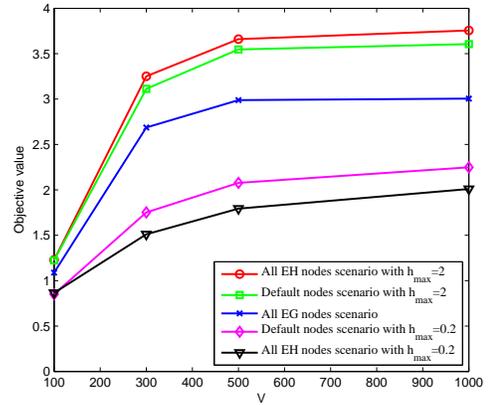}
\caption{The impact of node power supply manner and maximum available harvested energy on objective value.}
\label{fig:node_power_type}
\end{figure}

Next, we investigate the impacts of a variety of system parameters on the objective value, rate utility and electricity cost.
Fig. \ref{fig:node_power_type} shows the impact of the node power supply manner and  the maximum available harvested energy $h_{\max}$ on the objective value.
From Fig. \ref{fig:node_power_type}, we can see that
the lowest objective value is achieved at all EH nodes scenario with $h_{\max}=0.2$ much smaller than $g_{\max}=2$ and  the highest objective value is achieved at all EH nodes scenario with $h_{\max}=2$ equal to $g_{\max}=2$.
Due to the expense of the highest energy cost, all EG nodes scenario achieves the objective value lower than
all EH nodes scenario or default node scenario with $h_{\max}=2$.
In contrast, all EG nodes scenario achieves the objective value higher than
all EH nodes scenario or default node scenario with $h_{\max}=0.2$,
which results in the low energy supply and low data transmission.





\begin{figure}[t]
\centering
\includegraphics[width=2.8in]{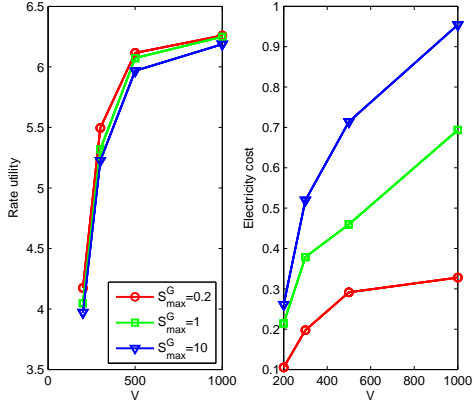}
\caption{The impact of different electricity prices on rate utility and energy cost.}
\label{fig:different_sgmax}
\end{figure}
 We investigate the impact of the electricity price on the rate utility and energy cost. We set three different electricity prices as $S_{\max}^G=0.2$, $S_{\max}^G=1$ and $S_{\max}^G=10$, respectively.
Fig. \ref{fig:different_sgmax} shows
that the electricity cost increases along with the increase of the electricity price. In order to reduce the electricity cost, EASYO reduce the energy consumption, and thus the corresponding rate utility decreases.

\begin{figure}[t]
\centering
\includegraphics[width=2.8in]{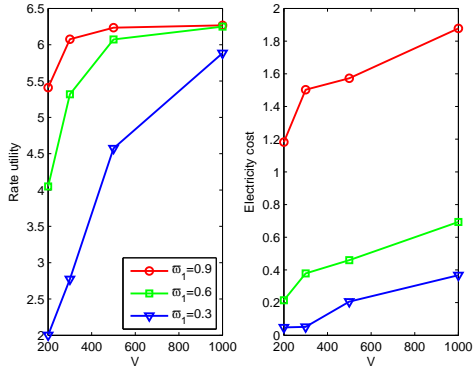}
\caption{The impact of different weight parameters on rate utility and energy cost.}
\label{fig:different_w1}
\end{figure}
 We investigate the impact of the weight parameter on the rate utility and energy cost. We set three the weight parameters as $\varpi_1=0.3$, $\varpi_1=0.6$ and $\varpi_1=0.9$,  respectively. When the weight parameter $\varpi_1$ is chosen as a large value, EASYO focuses on the rate utility maximization rather than the electricity cost minimization.
The results of Fig. \ref{fig:different_w1} verify this situation, where
the rate utility increases and the electricity cost also increases under a large value $\varpi_1=0.9$.

\begin{figure}[t]
\centering
\includegraphics[width=2.8in]{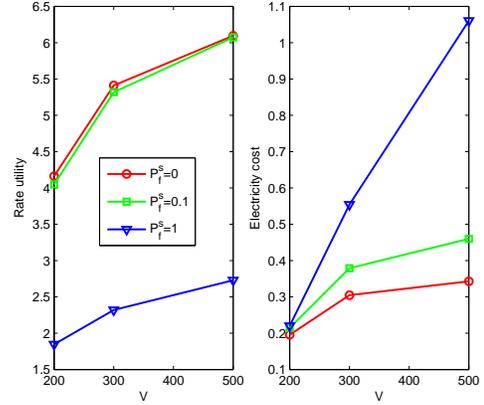}
\caption{The impact of different sensing energy consumption on rate utility and energy cost.}
\label{fig:SensingConsumption}
\end{figure}

Fig. \ref{fig:SensingConsumption} shows the impact of $\tilde{P}_f^S$ on the rate utility and energy cost. The larger $\tilde{P}_f^S$,
the more energy is required to supply for data sensing/processing, leading to
the less energy used in data transmission, and the lower rate utility.


\section{Conclusions}
Because of the instable energy supply and the limited battery capacity in EH node, it is very difficult to ensure the perpetual operation for WSN.
In this paper, we consider heterogeneous energy supplies from renewable energy and electricity grid, multiple energy consumptions and multi-dimension stochastic natures in the system model, and formulate a discrete-time stochastic cross-layer optimization problem to optimize the trade-off between the time-average rate utility and electricity cost.
To the end, we propose a fully distributed and
low-complexity cross-layer algorithm only requiring knowledge of the instantaneous system state.
The theoretic proof and the extensive simulation show that a parameter $V$ enables an explicit trade-off between the optimization objective and queue backlog.
In the future, we are interested in two aspects of delay
reduction by utilizing the shortest path concept, and by modifying the queueing disciplines.

\appendices
\section{Proof of Lemma 1}
\label{app:prop:P1}
Though squaring both sides of \eqref{data_queue}, we have (\ref{q}).
\begin{figure*}
\begin{eqnarray}\label{q}
&&\frac{1}{2}\left[ {{{\left( {Q_n^f(t{\rm{ + 1}})} \right)}^{\rm{2}}}{\rm{ - }}{{\left( {Q_n^f(t)} \right)}^{\rm{2}}}} \right]\nonumber\\
&& \le \frac{1}{2}{\left({{\textbf{1}}_{f \in {{\cal F}_n}}}{r_f}\left( t \right) + \sum\limits_{a \in {\cal I}\left( n \right)} {x_{{an}}^f\left( t \right)} \right)^{\rm{2}}}{\rm{ + }}\frac{1}{2}{{\rm{(}}\sum\limits_{b \in {\cal O}\left( n \right)} {x_{{nb}}^f\left( t \right)} )^2}\nonumber\\
&&{\rm{ + }}Q_n^f(t)\left({{\textbf{1}}_{f \in {{\cal F}_n}}}{r_f}\left( t \right) + \sum\limits_{a \in {\cal I}\left( n \right)} {x_{{an}}^f\left( t \right)}  - \sum\limits_{b \in {\cal O}\left( n \right)} {x_{{nb}}^f\left( t \right)} \right)\nonumber\\
&& \le {\left( {R_{\max }} \right)^{\rm{2}}}{\rm{ + }}\frac{3}{2}{\left( {{l_{\max }}{X_{\max }}} \right)^2}{\rm{ + }}Q_n^f(t)\left({{\textbf{1}}_{f \in {{\cal F}_n}}}{r_f}\left( t \right) + \sum\limits_{a \in {\cal I}\left( n \right)} {x_{{an}}^f\left( t \right)}  - \sum\limits_{b \in {\cal O}\left( n \right)} {x_{{nb}}^f\left( t \right)} \right)
\end{eqnarray}
\hrule
\end{figure*}
Similarly, we have (\ref{em}) from \eqref{EMqueue}.
\begin{figure*}
\begin{eqnarray}\label{em}
&&\frac{1}{2}\left[ {{{\left( {E_{n}\left( {t + 1} \right) - \theta _n^{{E}}} \right)}^2} - {{\left( {E_{n}\left( t \right) - \theta _n^{{E}}} \right)}^2}} \right]\nonumber\\
&& \le \frac{1}{2}\left({\left( {\textbf{1}_{n \in {{{\cal N}_H} \cup {{\cal N}_M}}}{e_n}\left( t \right) + \textbf{1}_{n \in {{{\cal N}_G} \cup {{\cal N}_M}}}{g_n}\left( t \right)} \right)^2+{\left( {p_n^{Total}\left( t \right)} \right)}^2}\right)\nonumber\\
&&+ \left( {E_{n}\left( t \right) - \theta _n^{{E}}} \right)\left({\textbf{1}_{n \in {{{\cal N}_H} \cup {{\cal N}_M}}}{e_n}\left( t \right) + \textbf{1}_{n \in {{{\cal N}_G} \cup {{\cal N}_M}}}{g_n}\left( t \right)-{{p_n^{Total}}}\left( t \right)}\right)\nonumber\\
&& \le \frac{1}{2}\left({\left( {\textbf{1}_{n \in {{{\cal N}_H} \cup {{\cal N}_M}}}h_{\max }+\textbf{1}_{n \in {{{\cal N}_G} \cup {{\cal N}_M}}}g_n^{\max} }\right)^2+\left( {p_{n,\max }^{Total}} \right)^2}\right) \nonumber\\
&&+ \left( {E_{n}\left( t \right) - \theta _n^{{E}}} \right)\left({\textbf{1}_{n \in {{{\cal N}_H} \cup {{\cal N}_M}}}{e_n}\left( t \right) + \textbf{1}_{n \in {{{\cal N}_G} \cup {{\cal N}_M}}}{g_n}\left( t \right)-{p_n^{Total}}\left( t \right)}\right)
\end{eqnarray}
\hrule
\end{figure*}
By plugging \eqref{q}, \eqref{em} and \eqref{Lyapunov_function} into \eqref{driftplus14}, we have \eqref{upperbound15A} with $B$ defined in \eqref{B_definition}.

\begin{figure*}
\begin{eqnarray}
{\widetilde{\Delta} _V}\left( t \right)&=&{ {\sum\limits_{n \in {\cal N}} {\sum\limits_{f \in {\cal F}} { Q_n^f(t)\left( {{\textbf{1}_{f \in {{\cal F}_n}}}{r_f}\left( t \right) + \sum\limits_{a \in {\cal I}\left( n \right)} {x _{{an}}^f\left( t \right)}  - \sum\limits_{b \in {\cal O}\left( n \right)} {{{x} _{{nb}}^f\left( t \right)} } }\right)} } } } \nonumber\\
&+& {\sum\limits_{n \in {{\cal N}}} {\left( {{{E_n}}\left( t \right) - \theta _n^{{E}}} \right)\left(\textbf{1}_{n \in {{{\cal N}_H} \cup {{\cal N}_M}}}{e_n}\left( t \right) + \textbf{1}_{n \in {{{\cal N}_G} \cup {{\cal N}_M}}}{g_n}\left( t \right) - p_n^{Total}\left( t \right)\right)} }\nonumber\\
 &-& V {\left( {\varpi _1}\sum\limits_{f \in {\cal F}} {{U_f}\left( {{{r}_f}(t)} \right)}-\left( {1 - {\varpi _1}} \right)\sum\limits_{n \in {{{\cal N}_G} \cup {{\cal N}_M}}} {\varpi _2}{P_n^G(t)}{{g_n}(t)}\right)}\label{upperbound15A}
  \end{eqnarray}
\hrule
\end{figure*}
Plugging  the definition \eqref{EnergyConsumption} of $p_n^{Total}\left( t \right)$ into \eqref{upperbound15A},
and rearranging all terms of the RHS in \eqref{upperbound15A}, ${\widetilde{\Delta} _V}\left( t \right)$ is changed into \eqref{upperbound15B}.
$\square$


\section{Proof of Part (A) in Theorem 1}
For $t=0$, we can easily have \eqref{Qbound}, then we assume \eqref{Qbound} is hold at time slot $t$, next we will show that it holds at $t+1$.

\textbf{Case 1}: If node $n$ doesn't receive any data at time $t$, we have
$Q_n^f\left( {t + 1} \right) \le Q_n^f\left( {t } \right) \le  {\varpi _1}{\beta_U}V + r_f^{\max }$.

\textbf{Case 2}: If node $n$ receives the endogenous data from other nodes
$a \in {\cal I}\left( n \right)$, we can get from \eqref{weightnformation} that
$W_{an}^f\left( t \right) - \sigma \ge 0$.
By plugging \eqref{link_information_weight} and \eqref{gamma}, we have $
{{Q}}_a^f\left( t \right) - Q_n^f\left( t \right)+A_{n}(t) \tilde P_n^R -\left( { r_f^{\max } + {l_{\max }}{X_{\max }}} \right )\ge0$.
Then, $Q_n^f\left( t \right) \le {{Q}}_a^f\left( t \right) -\left( { r_f^{\max } + {l_{\max }}{X_{\max }}} \right )+A_{n}(t) \tilde P_n^R $.
Due to $A_{n}(t)\leq 0$ and $\tilde P_n^R >0$, we have
  \begin{equation}\label{pf2}
    Q_n^f\left( t \right) \le {{Q}}_a^f\left( t \right) -\left( { r_f^{\max } + {l_{\max }}{X_{\max }}} \right ).
  \end{equation}
Plugging \eqref{Qbound} into \eqref{pf2}, we have
\begin{eqnarray}\label{add1}
  Q_n^f\left( t \right)&\le&  {\varpi _1}{\beta_U}V+r_f^{\max }- \left( { r_f^{\max } + {l_{\max }}{X_{\max }}}\right )\nonumber\\
  &=&{\varpi _1}{\beta_U}V-{l_{\max }}{X_{\max }}.
\end{eqnarray}
At every slot, the node can receive the amount of data at most $ r_f^{\max } + {l_{\max }}{X_{\max }} $. So
\begin{eqnarray} \label{add2}
  Q_n^f\left( {t + 1} \right) &\le& Q_n^f\left( t \right) + {l_{\max }}{X_{\max }} + r_f^{\max }.
\end{eqnarray}
Combing \eqref{add1} and \eqref{add2}, we have
$Q_n^f\left( {t + 1} \right) \le {\varpi _1}{\beta_U}V + r_f^{\max }$.

\textbf{Case 3}:
If node $n$ only receives the new local data,
according to \eqref{SourceRate}, the optimal value $r_f^*$ will met $
V{\varpi _1}U_f^{'}(r_f^*)=Q_n^f\left( t \right)  - A_{n}(t) \tilde P_f^S$, where $U_f^{'}(r_f(t))$ denotes the first derivative of $U_f(r_f(t))$. So we have $Q_n^f\left( t \right)  - A_{n}(t) \tilde P_f^S \le V{\varpi _1}{\beta_U}$, and then $Q_n^f\left( t \right)  \le {\varpi _1}{\beta_U}V$.
At every time, the new local data received at most is $r_f^{\max}$, so,
$ Q_n^f\left( {t + 1} \right) \le Q_n^f\left( t \right)  + r_f^{\max }\le{\varpi _1}{\beta_U}V+ r_f^{\max }$.

To sum up the above, we complete the proof of \eqref{Qbound}.

From \textbf{Remark 3.5} we can have \eqref{ehu}.$\square$

\section{Proof of Part (B) in Theorem 1}
The proof procedure of Part (B) in Theorem 1 is similar to that in \cite{Huang_Neely}, and hence is omitted for brevity.

\section{Proof of Part (C) in Theorem 1}
It is easy to verify that the following inequality holds according to the definition of ${C_{ab}}\left( {{{\bm p}^T}\left( t \right),{\bm S}\left( t \right)} \right)$:
\begin{equation}
 {C_{ab}}\left( {{{\bm p}^T}\left( t \right),{\bm S}\left( t \right)} \right) \le {C_{ab}}\left( {{{\bm p}^{T'}}\left( t \right),{\bm S}\left( t \right)} \right)\label{as2}
\end{equation}
where
${{\bm p}^{T'}}\left( t \right)$ obtained by setting
$p_{{nm}}^T\left( t \right)$ of
${{\bm p}^T}\left( t \right)$ to zero, $\left( {a,b} \right) \in {\cal L}$ and $\left( {a,b} \right) \ne \left( {n,m} \right)$.

For link $(n,m)$, plugging \eqref{link_information_weight} into \eqref{weightnformation}, we get
\begin{eqnarray}\label{addPartC0}
   \tilde W_{nm}^f\left( t \right) &=&{\left[ {Q_n^f(t)  - Q_m^f(t)+A_{m}(t) \tilde P_m^R - \sigma } \right]^ + }\nonumber\\
 &\le& {\left[ {Q_{n}^f\left( t \right) - \sigma } \right]^ + }
\end{eqnarray}

By plugging \eqref{gamma} and \eqref{Qbound} into \eqref{addPartC0}, we have
\begin{eqnarray}\label{addPartC1p}
 \tilde W_{nm}^f\left( t \right) &\le& {\left[ {{\varpi _1}{\beta_U}V + r_f^{\max } - {l_{\max }}{X_{\max }} - r_f^{\max }} \right]^ + }\nonumber\\
 &=&{\left[ {{\varpi _1}{\beta_U}V - {l_{\max }}{X_{\max }}} \right]^ + }
\end{eqnarray}
Since \eqref{addPartC1p} holds for any session $f$ through link $(n,m)$, we have
\begin{eqnarray}\label{addPartC1}
\tilde W_{nm}^{*}(t)&\le& {\left[ {{\varpi _1}{\beta_U}V - {l_{\max }}{X_{\max }}} \right]^ + }
\end{eqnarray}
We assume that
$E_{n}\left( t \right) < P_{n,\max }^{Total}$, when node
$n \in {{\cal N}}$
allocates nonzero power for data sensing, compression and transmission.
Furthermore, we assume that the power allocation control vector ${{\bm p}^{T*}}(t)$ is the optimal solution to \eqref{opeq}, and without loss of generality, there exists some $p_{{nm}}^{{T^{\rm{*}}}}(t) > 0$. By setting $p_{{nm}}^{T*} (t)= 0$ in ${{\bm p}^{T*}}(t)$, we get another power allocation control vector ${{\bm p}^T}(t)$.
We denote $G\left( {{{\bm p}^T}(t),{\bm S}(t)} \right)$ as the objective function of \eqref{opeq}. 
In this way, we get:
\begin{eqnarray}
&&\;\;\;\;G\left( {{{\bm p}^{T*}(t)},{\bm S}(t)} \right) - G\left( {{{\bm p}^T}(t),{\bm S}(t)} \right)\nonumber\\
&&= \sum\limits_{n \in {\cal N}} \sum\limits_{b \in {\cal O}\left( n \right)} [ {C_{nb}}({{\bm p}^{T*}(t)},{\bm S}(t)) \nonumber\\
&&- {C_{nb}}({{\bm p}^T}(t),{\bm S}(t))]\tilde W_{{nb}}^*\left( t \right)  \nonumber\\
&&+ \left ({E_{n}\left( t \right) - \theta _{n}^{E}}\right )p_{{nm}}^{T*}
\end{eqnarray}
From \eqref{as2}, we have
${C_{nb}}({{\bm p}^{T*}(t)},{\bm S}(t)) - {C_{nb}}({{\bm p}^T(t)},{\bm S}(t)) \le 0$ for
$b \ne m$. So
\begin{eqnarray}\label{addPartC3}
&&G\left( {{{\bm p}^{T*}(t)},{\bm S}(t)} \right) - G\left( {{{\bm p}^T(t)},{\bm S}(t)} \right)\nonumber\\
&&\le {C_{nm}}({{\bm p}^{T*}(t)},{\bm S}(t))\tilde W_{{nb}}^*\left( t \right)\nonumber\\
 &&+ \left ({E_{n}\left( t \right) - \theta _{n}^{E}}\right )p_{{nm}}^{T*}
\end{eqnarray}
According to our assumption $E_{n}\left( t \right) < P_{n,\max }^{Total}$ and the definition of $\theta_n^{E}$ in \eqref{thetaeh}, we have
\begin{eqnarray}\label{addPartC2}
  E_{n}^H\left( t \right) - \theta _n^{eH} &<& P_{n,\max }^{Total} - \theta _n^{eH}=- \delta {\varpi _1}{\beta_U}V
\end{eqnarray}
Plugging \eqref{PowerFeature1}, \eqref{addPartC1} and \eqref{addPartC2} into \eqref{addPartC3}, we have
\begin{eqnarray}
&&G\left( {{{\bm p}^{T*}},{\bm S}} \right) - G\left( {{{\bm p}^T},{\bm S}} \right)\nonumber\\
&&\le \delta p_{nm}^{T*}{\left[ {{\varpi _1}{\beta_U}V - {l_{\max }}{X_{\max }}} \right]^ + }- \delta {\varpi _1}{\beta_U}Vp_{{nm}}^{T*} \nonumber\\
&&< 0\nonumber
\end{eqnarray}

From the above inequalities, we can see that if $E_{n}\left( t \right) < P_{n,\max }^{Total}$, ${{\bm p}^{T*}}$  is not the optimal solution to \eqref{opeq}, which is opposite with our assumption. So, $E_{n}\left( t \right) \geq P_{n,\max }^{Total}$, which completes the proof of \eqref{ea1}.$\square$

\section{Proof of Part (D) in Theorem 1}

For node $n \in {{\cal N}}$, when any data of the $f$-th session is transmitted to other node, we can get $\tilde W_{nb}^f(t) > 0$.
From \eqref{weightnformation}, we have
${W_{nb}^f(t) - \sigma} > 0$.
By plugging \eqref{link_information_weight}, we have
$ Q_n^f(t)  - Q_b^f(t)+ A_{b}(t) \tilde P_b^R- \sigma > 0$.
Then, $Q_n^f(t)  >   \sigma +Q_b^f(t)- A_{b}(t) \tilde P_b^R$.
Since $Q_b^f(t)\leq 0$, $A_{b}(t)\leq 0$ and $\tilde P_b^R >0$, we have
 $Q_n^f(t) >  \sigma$.
By plugging the definition of $\sigma$ in \eqref{gamma},
we have $Q_n^f(t) >  {l_{\max }}{X_{\max }} + {r_f^{\max }}$, which completes the proof of \eqref{ea4}.

\end{document}